\documentstyle[epsfig,aps,preprint]{revtex}

\begin{document}
\draft
\title{Global Topology and Local Violation of Discrete 
Symmetries}
\author{J. Anandan}
\address{Raman Research Institute\\ C.V. Raman 
Avenue,Bangalore 560 080,
India}
\address{and}
\address{Department of Physics and Astronomy, University of 
South
Carolina,\\ Columbia, SC 29208, USA\\E-mail: jeeva@sc.edu}
\date{March 21, 97, Revised July 15, 97}
\maketitle

\begin{abstract}

Cosmological models that are locally consistent with general 
relativity and the standard model in which an object transported 
around the universe undergoes $P, C$ and $CP$ transformations, 
are constructed. This leads to generalization of the gauge fields 
that describe electro-weak and strong interactions by 
enlarging the gauge groups to include anti-unitary transformations.
Gedanken experiments show that if all
interactions obey Einstein causality then  P, C and CP cannot be 
violated in these models. But another model, which would violate charge 
superselection rule even for an isolated system, is allowed.
It is suggested that
the {\it fundamental} physical laws must have these discrete 
symmetries which are broken spontaneously,
or they must be non causal.
\end{abstract}
\vskip .5cm
\vskip .5cm
\pacs{hep-th/9706181, ~ PACS number: 01.55.+b, 98.80.H, 11.30.E }
\newpage
\par

The great success of the standard model has provided hardly any
experimental motivation to modify it at present. I
consider here some interesting physical consequences of 
generalizing the gravitational, electromagnetic, weak and strong 
fields, by modifying the global topology of an appropriate Kaluza-
Klein space-time. These generalizations are locally 
consistent with the {\it causal} dynamics of the standard model and
general 
relativity. Such topologies need to be studied also if in
quantum gravity the Feynman amplitudes for all 
possible topologies are summed. And yet, it will be shown by gedanken 
experiments
around global circuits in space-time that they are incompatible 
with the observed violations of parity ($P$), charge conjugation ($C$),
and $CP$ symmetries. This suggests that 
the standard model should be modified as will be discussed at the end.

Consider first a non orientable space. An example
is obtained by identifying a pair of opposite
faces of a rectangular box ( fig.~1)
continuously so that 
$A,B,C,D$
become identified with $A',B',C',D'$, respectively.
All sections parallel to $ABA'B'$ have the 
topology
of the Mobius strip $M^2$. Now let $AB, 
BC$ become infinite in
length while keeping $L=AB'$ large but finite. The
Cartesian product of this space with the real line $R$ is a non 
orientable manifold $M^4 = M^2 \times R^2$. This amounts to the 
identification
$(0,y,z,t)\leftrightarrow (L,y,-z,t)$. We
may take this manifold endowed with a flat Minkowskian metric to 
be
our space-time, denoted $S_1$. It trivially satisfies Einstein's 
field
equations in the absence of matter.  

In the presence of matter, we may consider the Einstein - de Sitter 
or the
Friedmann-Robertson-Walker cosmological model with zero 
spatial curvature,
with metric \cite{mtw} 
\begin{equation} 
ds^2\equiv
g_{\mu\nu}dx^\mu dx^\nu = -c^2 dt^2 + a^2(t) (dx^2 + dy^2 + dz^2). 
\end{equation} 
and
energy-momentum tensor 
\begin{equation} 
T^{\mu\nu} = (\rho
+P)u^\mu u^\nu + Pg^{\mu\nu}, 
\end{equation} 
where the density $\rho$ and
the pressure $P$ are constant in each hypersurface orthogonal to 
$u^\mu$. 
Then (1) and (2) satisfy the Einstein's field equations for 
appropriate
choices of $a(t),\rho(t)$ and $P(t)$ with $u^\mu=\delta^\mu_0$. 
For
example, for a pressure free universe (galaxies idealized as grains 
of dust with their random velocities neglected) $P=0$ and then
$a(t)=At^{2/3}$, whereas for radiation $P={1\over 3}\rho$ and 
then
$a(t)=Bt^{1/2}$, where $A$ and $B$ are constants \cite{mtw}, 
assuming zero
cosmological constant. All astrophysical evidence we have at present 
are
consistent with (1).
Again, this
cosmology may be made non orientable
by the identification described
above (fig. 1) to obtain a space-time, denoted $S_2$ \cite{ei1997}. 

Note that when the triad $OXYZ$ is
taken
to $O'$, identified with $O$, its $z-$axis has 
reversed direction compared to an 
identical
triad which was left at $O$, while the $X$ and $Y$ 
axes remain
the same (fig.~1). So, 
the triad has changed handedness around this closed curve. A left 
handed glove taken around any such closed curve, denoted 
$\Gamma$, will return as a right handed glove. Another interesting 
aspect of this and other space-times discussed here is 
that
they allow for some locally conserved quantities to be globally non
conserved. For example, the momentum 
$\bf p$ of a free
particle moving in the space $M^3=M^2 \times R$ described in fig. 
1 is {\it locally}
conserved, meaning that in any orientable neighborhood 
containing the particle $\bf p$ is conserved.
But if it goes around $\Gamma$ then its momentum 
component $p_z$ would 
have reversed.
Similarly, the angular momentum $\bf J$ of a 
torque free gyroscope would be 
locally conserved. Yet if it goes around 
$\Gamma$ then the $z-$ component $J_z$ of $\bf J$ 
would have
reversed. Therefore, $p_z$ and $J_z$ are not globally conserved.

To see how this is possible, note first that for every Killing field
$\xi^\mu_a$, the law $\nabla_\nu T^{\mu\nu} =0$ implies, via 
Killing's
equation, the local conservation law $\nabla_\mu j_a^\mu =0$ 
where $j_a^\mu
=T^{\mu\nu}\xi_{a\nu}$. The locally conserved momentum and 
angular momentum
components correspond to the independent translational and 
rotational
Killing fields of $M^3$. On defining the `charges' inside an 
oriented manifold $V$ with boundary 
$\partial V$
by $Q_a = \int_V \sqrt{-g}j_a^0 d^3x$, from Gauss' theorem,
\begin{equation} {dQ_a\over dt} = \int_{\partial V} \sqrt{-g}j_a^i 
dS_i . 
\end{equation} So, $Q_a$ is conserved iff the flux that is the RHS 
of (3)
vanishes. If $V$ is taken to be the interior of the box in fig. 1, then as 
a particle
goes out of $V$ through the end $A'B'C'D'$, because of the 
identification,
it simultaneously comes into $V$ through the opposite end 
$ABCD$ in $M^3$. Since
the identification reverses the $z-$ direction, it is
clear that the contribution that this simultaneous exit and entry 
the
particle makes to the RHS of (3)  vanishes for $Q_a = p_x, p_y, J_x, 
J_y$
but not for $Q_a = p_z, J_z$. Even the definitions of 
$p_z, J_z$ depend on the chosen neighborhoods; the above argument
shows that for the chosen maximal neighborhood they are not
conserved, unlike  $p_x, p_y, J_x$ and $J_y$.

The local $U(1)$ gauge symmetry of electromagnetism implies that 
the electromagnetic $U(1)$ group acts locally at each point in space-
time. This naturally leads to the 5 dimensional Kaluza-Klein (KK)
geometry that is obtained from space-time by replacing each point 
by a 
circle on which the $U(1)$ group acts locally, and conversely. The 
electromagnetic 
field
provides a $(1-1)$ correspondence between neighboring circles, 
called a
connection. As we go around a closed space-time curve, denoted 
$\gamma$, 
beginning and ending at a
point $o$, this correspondence leads to the rotation of the
circle at $o$, called the holonomy transformation of the
electromagnetic connection. When a wave function is taken around 
this curve
it is acted upon by this rotation and acquires the phase factor $\exp
(-ie\oint_\gamma A_\mu dx^\mu )$, which can be experimentally 
observed, where $A_\mu$ is the 
electromagnetic
potential. If $e$ is the smallest
unit of charge, then these phase factors for different closed curves
$\gamma$ define the electromagnetic field \cite{wu1975}.
Then, for a given $\gamma$ and electromagnetic field, these 
phase factors
may be used to define the
various charges that replace $e$.

The electromagnetic field may now be generalized by allowing the
identification of the $U(1)$ circles at $o$ to be in the opposite 
sense so that $\gamma$  is the projection of a 
Klein bottle in the KK space-time.
In particular, if $(x,y,z,t,\phi)$ are the coordinates of the KK space-
time, where $\phi$ is the angular variable in 
the fifth dimension, consider the slab of space-time $0\le x\le L$ 
with 
the identification of its ends by the homeomorphism 
$(0,y,z,t,\phi)\leftrightarrow (L,y,z,t,-\phi)$. Its projection on the 
usual space-time may be endowed with the metric (1). In this new KK 
space-time, denoted $S_3$, each two dimensional surface 
of constant $y,z,t$ is a Klein bottle $K^2$. So, 
$S_3$ is topologically $K^2\times R^3$.

Such a generalization amounts to enlarging the electromagnetic 
gauge group $U(1)$ to $O(2)$ that is generated by $SO(2) = 
U(1)$ and the reflection $E$ in two dimensional real Euclidean 
space. Although $O(2)$ is non abelian, because it is one 
dimensional, the gauge field is still  abelian. 

Suppose two observers start from the same point on $\gamma$ and 
go around $\gamma$ and meet. Each would then claim, 
with equal justification, that the charges of all the particles in the 
other observer have changed sign. So, it is not possible to determine 
unambiguously whether the sign of two charges at 
distinct points are the same. Because it is necessary to bring these 
charges to the  same space-time point in order to compare them, and 
the result would depend on the paths they take. Only the absolute 
value of the ratio of the charges would be meaningful.  Also, 
charge is locally conserved because of the 
$U(1)$ symmetry, but is not globally conserved. Because if two charges at
the same space-time point 
are taken along different paths and brought together again, their sum may 
change. This is 
similar to the global non conservation of momentum and angular 
momentum mentioned above, and can be understood in the same 
way by means of Gauss' theorem.

Also, suppose a charged particle wave function is split into two 
wave functions that are made to interfere around a closed curve 
having the property that the generalized electromagnetic holonomy
transformation associated with it is an improper $O(2)$ 
transformation. The superposed wave function then has the form
\begin{equation}
\psi (x^\mu,\phi) = exp(ie\phi)  \psi_1 (x^\mu) + exp(-ie\phi)  
\psi_2 (x^\mu),
\end{equation}
in a local gauge. 
Now, $\psi^*\psi$ has a 
non trivial $\phi$ dependence that makes it spontaneously break 
the $O(2)$ symmetry down to the discrete group consisting of $E$ 
and the identity. 
The charge
operator $Q=i{\partial\over \partial \phi}$.
Since $\psi$ is a superposition of opposite charges, it
violates the `charge superselection rule'.

This shows that the often made claim that the $U(1)$ gauge 
symmetry implies the charge superselection rule is incorrect, 
because the $O(2)$ gauge symmetry here contains $U(1)$. 
When Aharonov and Susskind \cite{ah1967} refuted this 
claim, they showed how a 
subsystem may be in a superpostion of charge eigenstates, while the 
entire system does not violate the charge superselection rule. An 
example is the BCS
ground state of a superconductor in which the Cooper pairs are in a 
superposition of different 
charge eigenstates, thereby breaking the electromagnetic $U(1)$ 
gauge symmetry spontaneously, 
while the entire superconductor may have 
a well defined charge and thus obey the charge superselection rule. 
But in the present case the entire system may 
be in superposition of charge eigenstates, and is therefore a stronger 
violation of this rule.

In the usual electromagnetic theory, $Q$ commutes with the {\it
interactions} so that the eigenstates of $Q$ form a `preferred basis' in
which the density matrix is diagonal. This gives an effective charge
superselection rule. In the present more general electromagnetic theory,
because the time evolution may contain $E$, which does not commute with
$Q$ that generates the electromagnetic $U(1)$, it is `easy' to produce a
superposition of opposite charges, as in the above example.  When such a
superposition interacts with an apparatus, the apparatus wave function
intensity also gets modulated correspondingly in the fifth dimension. {\it
This would make the fifth dimension observable}, like the other four
dimensions. 

This construction may be extended to the standard 
model for which the gauge group is $G= U(1)\times SU(2)\times
SU(3)$. 
The $C$ transform of a spinor
$\psi$ is $\psi^C = i\gamma^2 \psi^*$, where the
$*$ denotes complex conjugation or Hermitian conjugation in
quantum field theory. 
Therefore, as $\psi \rightarrow g\psi$ under $g\epsilon G$,
$\psi^C \rightarrow g^* \psi^C$. In the above 
construction each Klein bottle may be replaced by a
generalized Klein bottle that is closed by means of the 
automorphism $\alpha$ of $G$ that is the complex conjugation 
$\alpha (g) = g^*$ for every $g\epsilon G$, and 
the automorphism $\beta$ of the spinor Lorentz group
$\Lambda$ defined by $\beta (S) = 
i\gamma^2 S^* ( i\gamma^2)^{-1}= -\gamma^2 S^* \gamma^2$
for every $S\epsilon \Lambda$. If each fiber is a 
homogeneous space $G/H$, where $H$ is a subgroup of
$G$ such that $H^* = H$,
then the new KK space-time, denoted $S_4$, is
obtained by the identification 
$(0,y,z,t,Hg)\leftrightarrow (L,y,z,t,Hg^*)$. 
This performs a $C$
transformation on all the quantum numbers coupled to 
the gauge fields of $G$. 
Since the operational meaning of
a particle is contained in all its interactions, a 
particle taken around $\gamma$ in 
$S_4$ would become its anti-particle, 
i.e. it would undergo a $C$
transformation. For example, when taken around 
$\gamma$ a neutrino will return as an anti-neutrino. 

This leads to a generalization of the standard model in which the 
gauge group $G$ is enlarged to a group $\tilde G$ that is generated 
by $G$ acting on itself on the right and the automorphism 
$\alpha$. Then $\tilde G = O(2)\times \tilde{SU}(2)\times 
\tilde{SU}(3)$, where $\tilde{SU}(n)$ denotes the group of unitary 
and anti-unitary transformations on an $n$ dimensional complex vector
space that
have determinant $1$.
Even when $H$ is trivial so that $G/H = G$, 
$S_4$ is not a 
principal fiber bundle. Because the `twist' in the generalized Klein
bottle prevents the definition of the right action of $G$ 
everywhere. But $S_4$ is an associated bundle of a 
principal fiber bundle with structure group $\tilde G$ over the usual 
space-time as the common base manifold. Again by superposing $C$ 
eigenstates with distinct 
eigenvalues the extra dimensions become observable, as in the case 
of $S_3$ above.

A CP transformation may also be implemented physically by 
identifying the opposite faces of the slab according to 
$(0,y,z,t,Hg)\leftrightarrow (L,y,-z,t,Hg^*)$. This new KK space-time will 
be denoted by $S_5$. Time reversal may be implemented by the
identification $(0,y,z,t)\leftrightarrow (L,y,z,-t)$ 
so that space-time is 
time non orientable. But this would violate causality.

It is not necessary to go all the way around the universe to obtain 
the above discrete transformations. Cosmic strings, which 
are predicted to occur in the early universe, have been characterized 
by proper orthochronous Poincare transformations of the affine 
holonomy group around it \cite{to1994,an1996}. These solutions may 
be generalized to include also discrete transformations of 
the entire Poincare 
group as holonomy, e.g. reversal of the direction along the axis of the
cosmic string. The discrete holonomy
transformations would 
require taking out the axis of the cosmic string from space-time or 
turning it into a singularity. This would constitute a generalization of 
the gravitational field according to an earlier definition of the 
gravitational field \cite{an1996}. A generalized gauge field `flux' may 
also be introduced into the string by letting the gauge field holonomy 
around the string to include the new anti-unitary transformation 
$\alpha$ introduced above.

Except for $S_3$ (and its cosmic string analog) 
all the space-times discussed 
above are disallowed by the violation of discrete symmetries in weak 
interaction. In $S_1$ or $S_2$ consider two small capsules $U$ and 
$U'$ at the same location,
each containing the apparatus for the $P$ violating
experiment proposed by Lee and Yang \cite{le1956}, and 
performed by Wu et
al \cite{wu1957}. The magnetic coil, which orients the Co nuclei 
placed 
at the center of the coil, is in the x-y plane. When the 
nuclei
undergo $\beta$ decay, let the intensity distribution of
electrons be $f(\theta)$, where $\theta$ is the angle between the 
velocity
of the emitted electron and the $z-$ axis. Then,
$f(\theta)\ne f(180\deg -\theta)$, which violates $P$. Suppose 
now that the two capsules are taken along curves that form a circuit 
$\gamma$ such that the handedness changes during 
continuous transport around $\gamma$.
Let there be two twins in the
capsules 
performing the two respective experiments. When they meet again 
and compare their experiments they would find that the 
currents in
the two coils in the $X-Y$ plane are flowing in the same direction. 
However, 
the
distribution of the outgoing electrons
in $U'$ is $f(\theta ')=f(180\deg -\theta)$, which would be in conflict 
with the distribution
$f(\theta)$ obtained in the identical experiment performed in the 
capsule
$U$. Unlike the ``twin paradox'' in special relativity (which is not 
a paradox), here there is
perfect symmetry between the two twins: each twin would be 
justified in
saying that it is the other who has undergone a $P$ 
transformation. But the above contradiction disallows 
$S_1$ and $S_2$.

Since $\beta$-decay also violates $C$, $S_4$ is also disallowed by the 
above type of gedanken experiment. Similarly, $S_5$ is disallowed 
by doing identical experiments involving Kaon decay, which violates 
CP, in the two capsules. But $S_3$, with the generalized $O(2)$ 
electromagnetic gauge field introduced above, is consistent with 
all known phenomena. Because the charge reversal symmetry (or 
C restricted to purely electromagnetic phenomena) is an exact 
symmetry in all known 
phenomena. 

In an expanding universe, there may not be enough time 
for the capsules to go all the way around the universe and meet.
But in principle, we can set up a ring of large number
of capsules $\{U_1, U_2,...\}$ around the universe.
Two identical capsules $V_n$, $V_{n+1}$ 
meet midway between
two neighboring capsules $U_n$, 
$U_{n+1}$ 
at time $t=-T$; then $V_n$ meets $U_n$ and $V_{n+1}$
meets $U_{n+1}$ at $t=0>-T$. Finally,  $V_n$ and $V_{n+1}$
meet again at $t=T$ to verify if the 
relevant experiments in  $U_n$ and $U_{n+1}$
gave the same result at $t=0$. But in each of the above space-times,
except $S_3$, there would then be some $n$ for which the
experiments disagree, disallowing this space-time.

The restrictions due to the standard model on the global topology of 
the universe that it should not allow $P, C$ or $CP$ 
relative transformations around closed circuits is puzzling in view of 
the fact that the dynamics of the standard model is local and causal. I 
am requiring here that {\it any restriction on the boundary 
conditions 
must come from the laws of physics themselves}. This
raises the question of how the electroweak
interaction, which appears to be local and causal, could
influence the global topology of space-time or vice versa. As for 
the
possibility of the
former influence, as shown by the above examples of space-times, 
even Einstein's field equations do not in general determine the global
topology of space-time.
As for the possibility of the reverse influence, how could a
neutron `know' the global topology of space-time so 
that it
can safely decay in a $P$ violating way
without leading to the above contradiction if it were taken 
around
the universe? 

It appears that the simplest way of obtaining this connection is to 
suppose that
$ P, C$ or $CP$ is not violated by the laws of physics at the most 
fundamental
level, but that these symmetries are broken spontaneously. 
In the case of $P$ violation, there are then two sets of 
possible degenerate vacuua that are associated with the two possible
equivalent orientations. However, if the
boundary conditions in the early universe are such that space is 
non
orientable then neither vacuum can be chosen all the way around 
the
universe because this would result in a mismatch. So, the
reflection symmetry is either not spontaneously broken, or
broken in orientable domains in which the
$P$ violation may be different corresponding to the two different
possible orientations. But if the boundary
conditions are such that space is orientable then a vacuum with the 
same orientation may be chosen everywhere so that $P$ is violated
in the same manner.
It is emphasized that the spontaneous breaking of symmetry 
may occur due
to local, causal physics. 

Another possiblility is to give up Einstein causality in the 
fundamental
physical laws, which was assumed to obtain the above 
contradictions. Since
this principle is very well confirmed by all our experience, it 
appears
that we could give it up only \cite{tachyon} in 
the early universe when
quantum gravitational effects were important. This
appears reasonable also because quantum gravity requires the
quantization of space-time geometry including its causal structure, 
and is
therefore inherently non causal and perhaps also non local. So, if 
quantum gravity violates  $P, C$ and $CP$ intrinsically, 
then its laws
could determine, in a non causal way, the 
global
topology of the KK space-time to be compatible with these violations. 
Then the  $P, C$ and $CP$ violating 
structure of
the standard model would be obtained in the low energy limit of 
such a non
causal quantum gravity.

As for the first of the above two possibilities, 
purely left-right symmetric models with spontaneous violation of P 
have 
been proposed \cite{se1975}. But the above argument implies that, 
in this approach, $C$ and $CP$ should also be violated spontaneously. The 
second possibility, above, requires quantum gravity 
to be
non causal and $P,C,CP$ violating, and would therefore heuristically 
guide
us in the construction of a quantum theory of gravity. It has the
advantage over the first possibility that the non causal nature 
could
resolve another puzzle. This is the horizon
problem \cite{mi1968}, namely the
fact that regions in the early universe which are causally 
unrelated
nevertheless have similar properties, such as temperature and 
density,
which appears to violate Einstein causality.

If quantum gravity, which is expected to unify all the interactions, 
were
to violate $P, T, C$ and $CP$, then it is not surprising that 
the electro-weak theory
obtained as a low energy limit of quantum gravity should 
also
violate these symmetries. It would equally not be surprising if the
classical
gravitational field, which is also a low energy limit of quantum 
gravity,
should contain a residual violation of $P, C$ and $CP$ ( $C$ and $CP$ are the 
same as  $PT$ 
and $T$, respectively, if we assume $CPT$ symmetry). It is therefore 
worthwhile to
look for experimental evidence of violation of these discrete 
symmetries in the 
gravitational
interaction \cite{da1976,an1982}. 

I thank Yakir Aharonov, Ralph Howard, Simon Donaldson and Paule 
Saksidas for useful
discussions. This work was supported by NSF grant PHY-9601280.

\newpage
\begin{figure}
\epsfig{file=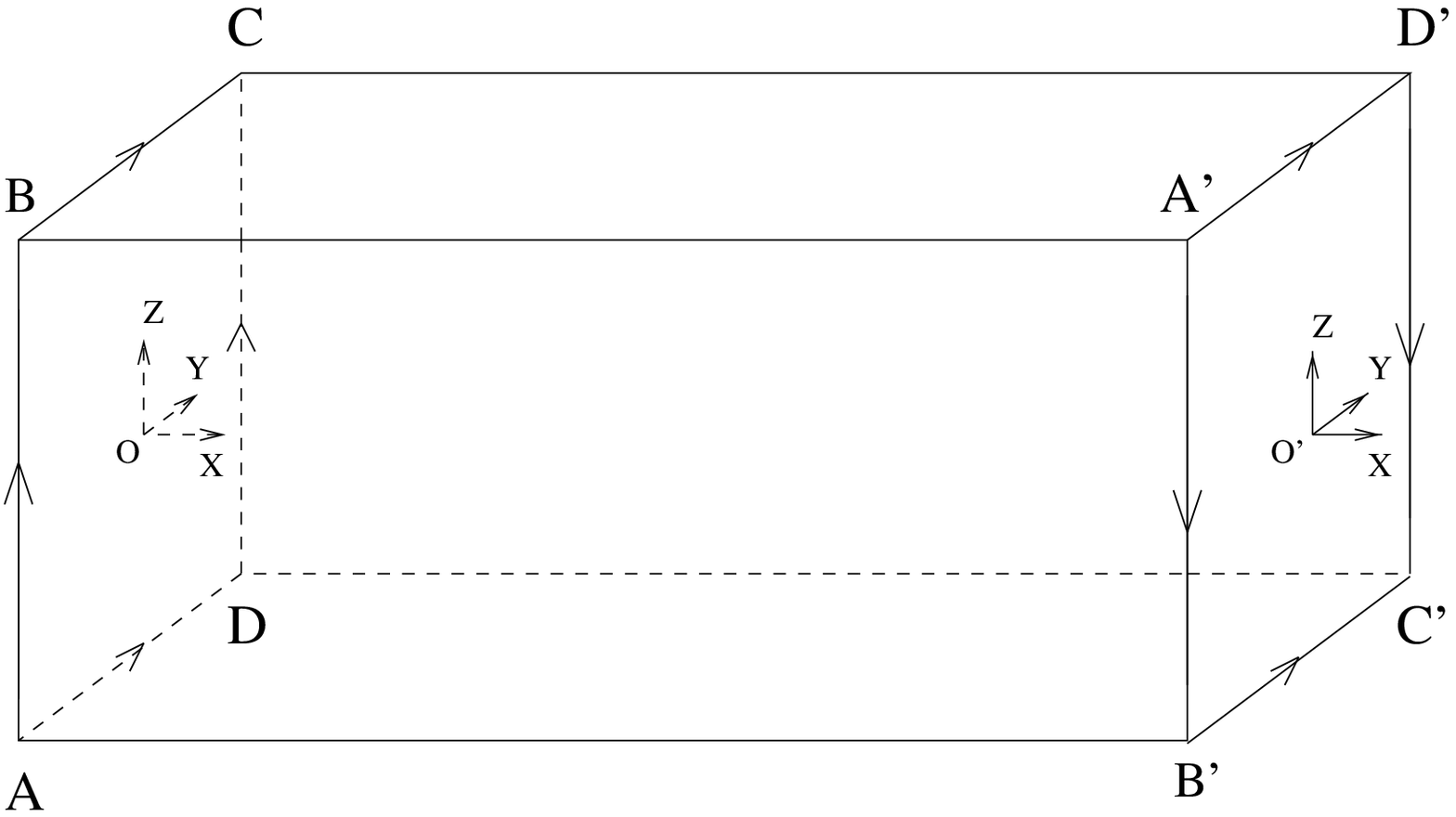,height=6cm}
\vskip .5cm
\noindent
\caption[]{A non orientable space with zero spatial curvature, 
obtained 
by identifying the ends $ABCD$ and $A'B'C'D'$ so that each primed 
letter
represents the same location as the unprimed one.} 
\end{figure}


\begin{thebibliography}{99}
\bibitem{tachyon} An alternative way of having a non causal, 
although
local, interaction is
by means of tachyons. However, tachyons would make the 
vacuum unstable and
therefore do not appear to be a realistic alternative.
\end{thebibliography}
\end{document}